\documentstyle[12pt]{article}
\begin{document}
\rightline{YPI 1407 (18)-93}
\rightline{BONN-HE-93-30}
\rightline{hep-th/9310110}
\vspace{1cm}
\begin{center}
{\Large Off-Shell Bethe Ansatz Equation for Gaudin Magnets and Solutions
of Knizhnik-Zamolodchikov Equations}
\end{center}
\vspace{2cm}
\begin{center}
H.M.\ Babujian{$^{\ast \dag}$} and R.\ Flume{$^{\ddag}$}
\end{center}
\vspace{2cm}
{\bf Abstract} \\
We generalize the previously established connection between the off-shell
Bethe ansatz equation for inhomogeneous $SU(2)$ lattice vertex models in the
quasiclassical limit and the solutions of the $SU(2)$ Knizhnik-Zamolodchikov
equations to the case of arbitrary simple Lie algebras.
\vspace{4cm} \\
{}$^{\ast}$ Yerevan Physics Institute, Alikhanian Brothers 2,
Yerevan, 375036 Armenia \\
{}$^{\dag}$ Partially supported by the grant 211-5291 YPI of the German
Bundesmini\-sterium f\"ur Forschung und Technologie \\
{$^{\ddag}$} Physikalisches Institut der Universit\"at Bonn, Nussallee 12,
D-53115 Bonn, Germany

\newpage
\section{}
One of the present authors has derived some time ago, [1], a connection (in
fact an isomorphism) between the two problems, the diagonalization of Gaudin
type magnetic Hamiltonians [2] and the realization of solutions of rational
Knizhnik-Zamolodchikov equations (KZE) [3]. The deductions in [1] were
performed with $SU(2)$ as underlying symmetry algebra. A streamlined version
of the argument together with the demonstration that the eigenvectors of the
Gaudin Hamiltonian can be viewed as quasiclassical approximation of the
Bethe wave vectors of an inhomogenous vertex model, can be found in [4].

We briefly review the essence of the reasoning from [1] and [4]. Let
denote by $T(\mu \{ Z_i \})$ the transfer matrix of an integrable vertex
model, where $\mu$ denotes a global
spectral parameter and $Z_1 , \ldots , Z_N$ are local disorder parameters.
$T(\mu \{Z_i\})$ acts on a n-fold tensor product of $SU(2)$ representation
spaces. The technique of the algebraic Bethe ansatz renders a vector
$\phi (\mu , \ldots , \mu_m )$ in the tensor space satisfying by
construction the following equation [5]
\[
T( \mu \{ Z_i \} ) \phi ( \mu_1 , \ldots , \mu_m , \{ Z_i \} ) =
\Lambda ( \mu_1 , \ldots , \mu_m , \{Z_i \} ) \phi (\mu_1 , \ldots ,
\mu_m \{Z_i \} )
\]
\begin{equation}
- \sum^{m}_{\alpha = 1} \frac{F_\alpha \Phi_\alpha}{\mu - \mu_\alpha}
\end{equation}
where $F_\alpha (\mu_1 , \ldots , \mu_m , \{Z_i \})$ denote some
c-number functions and $\Phi_\alpha$ represent some vectors which are not
parallel to $\phi$. Following the lines of the Bethe ansatz one chooses
the parameters $\mu_1 , \ldots , \mu_m$ such that the
$F_\alpha (\mu_1 , \ldots , \mu_m , \{Z_i \} )$ vanish. Eq. (1) becomes then
a true eigenvalue equation with eigenvalue
$\Lambda ( \mu_1 , \ldots , \mu_m ,\{ Z_i \}) $. This is however
not the route
to be followed here driving to a connection with the KZE. Instead one keeps
$\mu_1 \ldots \mu_m$ in general position. Eq. (1) was termed in [4] the
Off-Shell Bethe ansatz Equation (OSBAE). The main result of [1, 4] is that
suitable integrals in $\mu_1 ,\ldots , \mu_m$ of
$\phi (\mu_1 , \ldots , \mu_m , \{Z_i \})$,
(evaluated in quasiclassical approximation), together with an integrating
factor give rise to a solution vector of the KZE. \\
The (presumably) complete solution of the vector-valued rational KZE is
due to Schechtman and Varchenko [6] (see also Matsuo [7]). An elegant
rederivation of the Schechtman-Varchenko solution has been given by Awata,
Tsuchiya and Yamada [8]. Using the
result of these authors we shall establish here the above mentioned
connection between Gaudin Hamiltonians and the KZE for arbitrary simple
Lie algebras. The gain will be, as we think, twofold.
The first achievement is a simplified
verification of KZE solutions. Secondly, and more important, one obtains an
explicit realization of wave vectors of Bethe ansatz problems (in the
quasiclassical limit) which were previously only represented implicitely
through an hierarchical imbedding procedure [9, 10, 11]. It remains as a
challenge to find a similary universal representation of the full
(non semiclassical) Bethe ansatz problem.
\section{}
Let $E_{\pm \alpha}$ and $H_\alpha$ denote the generators of a simple
Lie algebra corresponding to simple root vectors $\alpha$ and to
generators of the Cartan subalgebra respectively with commutation relations
\begin{equation}
\left.
\begin{array}{lcl}
\left[ E_\alpha , E_{- \alpha} \right]           & = & h_\alpha \\
\left[ E_\alpha , E_\beta \right] & = & N_{\alpha , \beta}
                        E_{\alpha + \beta} \\
\left[ h_\alpha , E_\beta \right] & = & (\alpha , \beta ) E_\beta
\end{array} \right\}
\end{equation}
where $N_{\alpha,\beta}$ denote some constants and $(\alpha,\beta )$ refers
to some inner product in the space of root vectors. (A complete description
of the Lie algebra includes the specification of Serre relations which we
leave aside since we don't use them in the following). We consider vector
spaces $V^a$ carrying an irreducible representation of the algebra (2),
characterized through a highest weight vector $\lambda^a$. With
$E^a_\alpha, h^a_\alpha$ etc. will be denoted the representatives of the
Lie algebra generators on $V^a$. We will make use of the operators
\begin{equation}
\left.
\begin{array}{lcl}
E_{\pm \alpha} (\mu)
& = &
\sum^{N}_{a=1} \frac{E^a_{\pm \alpha}}{\mu - Z_a} \\ \\
h_{\alpha} (\mu)
& = &
\sum^{N}_{a=1} \frac{h^a_{\alpha}}{\mu - Z_a}
\end{array}
\right\}
\end{equation}
acting on the tensor space $\Omega = V^1 \otimes V^2 \otimes \ldots
\otimes V^N$. The $Z_a , a = 1, \ldots , N$ are here coordinates on
$C^N$. \\
The KZE [12] for vector-valued functions $\psi (Z_1 , \ldots , Z_N) \in
\Omega$ reads as
\begin{equation}
\kappa \frac{\partial \psi}{\partial Z_a} = \sum_{b \neq a}
\frac{(t^a \cdot t^b )}{Z_a - Z_b} \psi
\end{equation}
with $(t^a \cdot t^b )$ denoting the Killing form with respect to the Lie
algebra generators in the spaces $V^a$ and $V^b$. The parameter $\kappa$
is (for our purposes) an arbitrary non zero complex number. To quote the
general solution, due to Schechtman and Varchenko [6], of this equations we
need some more notations. The monodromy of the solutions is determined by a
scalar function $\chi ( \mu , \ldots , \mu_m , Z_1 , \ldots , Z_N)$ of the
form
\begin{equation}
\chi = \prod^m_{i < j}
         (\mu_i - \mu_j )^{\frac{( \alpha_i , \alpha_j)}{\kappa}}
\prod^N_{a < b} (Z_a - Z_b )^{\frac{( \lambda^a , \lambda^l)}{\kappa}}
\prod^{m, N}_{i, a} (\mu_i - Z_a)^{\frac{(\alpha_i , \lambda^a )}{\kappa}}.
\end{equation}
$(\lambda^a , \lambda^b )$ and $(\alpha_i , \lambda^a )$ stand for the
scalar product introduced in Eq. (2). The solution $\psi (Z_1 , \ldots
, Z_N )$ of the KZE (4) is represented by a multiple contour integral in
the complex variables $\mu_1 , \ldots , \mu_m$ of the product of the
function $\chi$ with a vector function $\varphi (\mu, Z)$, as follows
\begin{equation}
\psi (Z_1 , \ldots , Z_N ) = \oint \ldots \oint \chi (\mu , Z )
 \varphi ( \mu , Z) d \mu , \ldots , d \mu_m .
\end{equation}
The vector $\varphi ( \mu , Z )$ can be determined through a recursive
procedure in the number of variables $\mu$:
\[
\varphi (\mu_1 , \ldots , \mu_m ) = X^m (\mu_1 , \ldots , \mu_m ) \omega
\]
with $\omega$ being the vector of highest weight in
$\Omega , h^a_\alpha \omega = \lambda^a \omega , \; E^a_\alpha \omega = 0 ,
a = 1 \ldots N , \omega \equiv \prod_a | \lambda^a >$,
$$
\begin{array}{lcl}
X^1 (\mu_1 ) & = & E_{-\alpha_1} (\mu_1 ) ,  \\
X^m (\mu_1 , \ldots , \mu_m ) & = & X^{m-1} (\mu_1 , \ldots ,
\mu_{m-1}) X^1 (\mu_m ) \\
& &       + D(\mu_m ) X^{m-1} ( \mu_1 , \ldots , \mu_{m-1}) ,
\hspace{3.8cm}  ({\rm 7})\\
X^{m-1} & = & X^{m-1} (E_{- \alpha_1} , \ldots , E_{- \alpha_m}) \\
D (\mu_m ) X^{m-1} & = & \sum\limits^{m-1}_{i = 1} \frac{1}{\mu_m - \mu_i}
       X^{m-1} \left( E_{- \alpha_1}, \ldots ,
\left[ E_{- \alpha_m} , E_{- \alpha_i} \right]
\ldots E_{- \alpha_m} \right) .
\end{array}
$$
The recursion relation (7) has the solution
\setcounter{equation}{7}
\begin{equation}
\varphi = \sum_{\mbox{\scriptsize{part}}} \prod^N_{a = 1}
\sum_{\mbox{\scriptsize{perm}} \atop \alpha_{i_a}\in I_a}
\frac{E^a_{- \alpha_{1_a}} \cdot E^a_{- \alpha_2} \ldots
E^a_{- \alpha_{q_a}} | \lambda^a >}
{(\mu_{1_a} - \mu_{2_a} ) (\mu_{2_a} - \mu_{3_a} ) \ldots (\mu_{q_a} - Z_a )}
\end{equation}
where the first sum is over N-fold partitions
\[ \left( I_1 , \ldots , I_N ,
\bigcup\limits^N_{a=1} I_a = ( \alpha_1 , \ldots , \alpha_m) ,
I_a \cap I_b = \emptyset \; a \neq b \right) \]
of the of group labels
$\alpha_1 , \ldots , \alpha_m$ (empty subsets are included) and the
second sum is over permutations in a given subset. One had to fix
the contours of the m-fold integral in $\mu_1 , \ldots , \mu_m$ for
a complete specification of the solution vector $\psi$, Eq. (6).
We will not go into the (involved) argument for an appropriate choice
of a basis of contours but only note that the contours have to be
taken as closed so that integrals along them of functions,
representable as total derivatives in $\mu_1 , \ldots , \mu_m$, vanish.
Finally, we remark that $\psi$ is highest weight vector in $\Omega$
of the heighest weight
\[
\lambda_\psi = \sum^N_{a = 1} \lambda^a - \sum^m_{i = 1} \alpha_i
\]
\section{}
To start with the discussion of Gaudin type Hamiltonians we briefly sketch
their a appearance in the quasiclassical limit of integrable vertex models
[2, 11, 13, 4]. Let $R^{0j} (\mu , \eta)$ and $R^{00} (\mu , \eta )$ resp.
denote solutions of the Yang-Baxter equation attached to some simple Lie
algebra. $\mu$ is the usual spectral parameter whereas $\eta$ describes the
distance of $R$ from its classical limit
\begin{equation}
\left.
\begin{array}{lcl}
R^{00} (\mu , \eta = 0 ) & = & I^0 \otimes I^0 \\
R^{0j} (\mu , \eta = 0 ) & = & I^0 \otimes I^j
\end{array}
\right\}
\end{equation}
$R^{00} ( \mu , \eta )$ and $R^{0j} (\mu , \eta )$ act on tensor products
$V^0 \otimes V^0$ and $V^0 \otimes V^j$ respectively, where $V^0$ carries
some appropriately chosen basic representation and $V^j$ an arbitrary
representation of the group under consideration. $I^0 , I^i$ in Eq. (9)
denote the unit operators in $V^0$ and $V^j$ respectively. An "appropriate"
choice of a basic representation means that one satisfies through it the
Yang-Baxter algebra relation
\begin{equation}
R^{00} (\mu - u ) R^{0j} (\mu )R^{0i} (u) = R^{0i} (u) R^{0j} (\mu )
 R^{00} (\mu - u) .
\end{equation}
Eq. (11) allows the introduction of a family of commuting transfer matrices
acting on the tensor product space $\Omega = \prod^N_{a = 1} \otimes V^a$.

One achieves this [5] , by first introducing a monodromy operator
\begin{equation}
J (\{ \mu_i \} , \{ Z_a \} ) = \prod^N_{a = 1} R^{0 a} (\mu - Z_a )
\end{equation}
where the global spectral parameter $\mu$ has been affiliated with local
disorder parameters $Z_1 , \ldots , Z_N$. The transfer matrices are given as
\begin{equation}
T ( \mu , \{ Z_a \} ) = t_{r_0} J ( \mu , \{ Z_a \} )
\end{equation}
with $t_{r_0}$ denoting the trace with respect to the space $V^0$.
The $R$-matrices $R^{00}, R^{0j}$ have in the rational case at hand the
quasiclassical expansion.
\begin{equation}
\left.
\begin{array}{lcl}
R^{00} ( \mu - Z_a ) & = & I^0 \otimes I^0 +
                  \frac{\eta}{\mu - Z_a} (t_0 \cdot t_0 ) +  0(\eta^2 ) \\
R^{0i} ( \mu - Z_a ) & = & I^0 \otimes I^i +
                  \frac{\eta}{\mu - Z_a} (t_0 \cdot t_i ) +  0(\eta^2 )
\end{array}
\right\}
\end{equation}
with the notation as in Eq. (5). The second term on r.h.s. of (13) is a
solution of classical Yang-Baxter equation [14]. Inserting (13) into
Eq. (12) we arrive at
\begin{equation}
T ( \mu , \{ Z_i \} ) = {\mbox{const}} \cdot I + \eta^2
                        \sum^N_{j = 1} \frac{H_a}{\mu - Z_a} + 0 ( \eta^3 )
\end{equation}
\begin{equation}
H_a = \sum^N_{b \neq a} \frac{(t_a \cdot t_b )}{Z_a - Z_b} .
\end{equation}
The linear term in $\eta$ drops out for simple Lie algebras.

The diagonalization of $T ( \mu , \{ Z_i \} )$ and a fortiori of its
quasiclassical
limit (14) can in principle be obtained, as observed by several authors
[4, 10, 11] with help of the nested algebraic Bethe ansatz technique [9].
The first step of this procedure leads, as mentioned above, to a vector
$\Phi$ satisfying the OSBAE
\[
T( \mu , \{ Z_i \} ) \Phi ( \mu_1 , \ldots , \mu_m, \{ Z_i \} )  =
\Lambda ( \mu_2 , \ldots , \mu_m , \{ Z_i \} ) \Phi ( \mu_1, \ldots ,
 \mu_m , \{ Z_i \} ) - \]
\[
 - \sum^m_{i = 1} \frac{F_i \Phi_i}{\mu - \mu_i} .
\]
The problem with this approach is that it is not known up to now how to
resolve the nested Bethe ansatz explicitely for the wave vector. We can
however circumvent this deadlock (in the quasiclassical approximation)
by the use of the
Schechtman-Varchenko (SV) result and capitalizing on the experience gained
with the $SU(2)$ vertex model. We claim in short that the vector $\varphi$
appearing in the integrand of the SV solution, Eq. (7), satisfies the
quasiclassical version of the OSBAE,
\begin{equation}
H_a \varphi = h_a \varphi + \sum^m_{j = 1} f_{\mu_j}
\overline{\varphi}^a_{\mu_j}
\end{equation}
where $h_a$, $f_{\mu_j}$ and $\overline{\varphi}^a_{\mu_j}$ are given by
\begin{equation}
h_a = \sum^N_{b \neq a} \frac{( \lambda^a , \lambda^b )}{Z_a - Z_b} -
      \sum^m_{j = i}    \frac{( \alpha_j , \lambda^a )}{Z_a - \mu_j}
\end{equation}
\begin{equation}
f_{\mu_j} = \sum^m_{i \neq j} \frac{(\alpha_j , \alpha_i )}{\mu_j - \mu_i} -
            \sum^N_{a = 1}    \frac{(\alpha_j , \lambda^a)}{\mu_j - Z_a}
\end{equation}
\begin{equation}
\overline{\varphi}^a_{\mu_j} =
\sum_{\mbox{\scriptsize{part}} \atop \alpha_j \in I_a}
\prod^N_{b = 1} \sum_{\mbox{\scriptsize{perm}}}
\frac{E^b_{- \alpha_{1_a}} E^b_{- \alpha_{2_a}} \ldots
  E^b_{- \alpha_{q_a}} | \lambda^b >}
 {(\mu_{1_a} - \mu_{2_a}) (\mu_{2_a} - \mu_{3_a}) \ldots (\mu_{q_a} - Z_q )}.
\end{equation}
The last expression is equal to the subset of terms on the r.h.s. of Eq. (9)
with $\alpha_j \in I_a$. Eqs. (16-19) can be proven inductively with the
number $m$ as induction parameter. The verification on the level $m = 2$
will be given in the next section. This calculation already displays the
main features of the
general inductive argument which will be presented in the forthcoming paper
[14]. We proceed for now with the demonstration that the KZE is indeed a
consequence of the OSBAE(16). To start with we introduce a function $\chi$
by the differential equations
\begin{equation}
\kappa \frac{\partial \chi}{\partial Z_a} = h_a \chi
\end{equation}
\begin{equation}
\kappa \frac{\partial \chi}{\partial \mu_j} = f_{\mu_j} \chi .
\end{equation}
Eqs. (20-21) are easily shown to be compatible and are integrated to the
result written down in (6). The point in introducing $\chi$ anew through
the relations (20-21)  is that we want to emphasize that $\chi$ plays the
role of an integrating factor in what follows. One may in fact motivate
the introduction of $\chi$ through the requirement alone that the
differential Z-dependence of the product $\chi \varphi$ reproduces modulo
a total differential in the variables $\mu$ the r.h.s. of the OSBAE. The
very possibility to find ab initio the integrating factor starting from
the OSBAE is of cours anchored in the structure of the vectors of the
OSBAE itself. We think that the neat interrelation between the
behaviour of the wave vectors with respect to the disorder parameter
$Z_1 , \ldots , Z_N$ and the spectral parameters $\mu_1 , \ldots , \mu_m$
should be considered as key point of the approach followed in this paper
and in ref. [4] and [1]. Straightforward differentiation gives
\begin{equation}
\frac{\partial \varphi}{\partial Z_a} =
\sum^m_{i = 1} \frac{d}{d \mu_i}
\sum_{\mbox{\scriptsize{part}} \atop \alpha_1 \in I_a}
\prod^N_{b = 1}
\sum_{\mbox{\scriptsize{perm}}}
\frac{E^b_{- \alpha_{1_b}} \ldots E^b_{- \alpha_{q_b}}}
{( \mu_{1_b} - \mu_{2_b} ) \ldots ( \mu_{q_b} - Z_b )} | \lambda^b > .
\end{equation}
We obtain from Eqs. (20-22)
\begin{equation}
\frac{\partial}{\partial Z_a} ( \chi \varphi ) =
h_a \chi \varphi + \sum^m_{j = 1} \chi f^a_{\mu_j} \varphi^a_{\mu_j} -
\sum^m_{j = 1} \frac{\partial}{\partial \mu_j} (\chi \varphi^a_{\mu_j})
\end{equation}
{}From the OSBAE (16) and Eq. (23) follows immediately that
$\psi (Z_1 , \ldots , Z_N )$, Eq. (6), satisfies the KZE.
\section{}
We finally present the detailed verification of OSBAE for $m = 2$.
To prepare the verification of the OSBAE for $m = 2$ we introduce
\begin{eqnarray}
C(\Lambda ) & = & \left( t (\Lambda ) \cdot t ( \Lambda ) \right) ,
       \nonumber \\
t (\Lambda ) & = & \sum^N_{a = 1} \frac{t^a}{\Lambda - Z_a} ,
\end{eqnarray}
where we use notations as in Eq's. (4) and (5). The desired results for
$H^a$ are a simple consequence of the corresponding results for
$C( \Lambda )$ because one has for $\Lambda \approx Z_a$ the expansion
\[
c (\Lambda ) = \frac{1}{( \Lambda - Z_a )^2}c^a_2 + \frac{2}{\Lambda - Z_a}
H^a + O \left( ( \Lambda - Z_a)^0 \right).
\]
$c^a_2$ is here the quadratic Casimir operator on $V^a$.

One finds with the Lie algebra relations (2) the following identity
\begin{eqnarray}
\left[ C(\Lambda ) , E_{- \alpha}(\mu) \right]
& = &
\{ E_{-\alpha} (\Lambda) h_\alpha (\mu) -
E_{- \alpha} (\mu) h_\alpha (\Lambda ) +  \nonumber \\
& &
+ \sum_{\beta > 0 , \beta \neq \alpha} N_{- \alpha , - \beta}
\left( - E_{- \alpha - \beta } (\Lambda )
E_\beta (\mu) \right. \nonumber \\
& & \left. \left. + E_{- \alpha - \beta} (\mu)
   E_\beta (\Lambda) \right) \right\}
\end{eqnarray}
{}From the last equation follows that
\[
\varphi (\mu) \equiv X^1 (\mu) \omega \equiv E_{- \alpha} (\mu) \omega
\]
satisfies the OSBAE (16) with $m = 1$. In fact, exploiting Eq. (25)
we obtain:
\begin{equation}
\begin{array}{lcl}
C( \Lambda ) E_{- \alpha} (\mu) \omega & = & \frac{2}{\Lambda - \mu}
\left( E_{- \alpha} ( \Lambda ) h_\alpha (\mu) \right.\\
& & \left.- E_{- \alpha} (\mu) h_\alpha ( \Lambda ) \right) \omega +
E_{- \alpha} ( \Lambda ) C( \Lambda ) \omega
\end{array}
\end{equation}
The second term $\sum_{\beta >} ( \ldots )$, on the r.h.s. of Eq. (25)
does not contribute here as it produces a vanishing result acting on the
highest weight vector $\omega$. Evaluating $h_\alpha (\mu) ,
h_\alpha ( \Lambda )$ and
$C (\Lambda )$ resp. as applied to $\omega$ one arrives after restriction
to the pole term at $\Lambda = Z_a$ immediately at the OSBAE, Eq. (16). \\
The quasiclassical Bethe wave vector $\varphi (\mu_1 , \mu_2 )$, Eq. (7),
reads as
\begin{eqnarray}
\varphi (\mu_1 , \mu_2 ) & = & X^2 (\mu_1 , \mu_2 ) \omega \nonumber \\
   X^2(\mu_1 , \mu_2 ) & = & E_{- \alpha_1} (\mu_1) E_{- \alpha_2}
    (\mu_2 ) +
D (\mu_2 ) (E_{- \alpha_1}(\mu_1 ))
\end{eqnarray}
The verification of the OSBAE for $\varphi (\mu_1 , \mu_2 )$ is again a
matter of direct calculation. We have
\begin{eqnarray*}
C(\Lambda ) \varphi ( \mu_1 , \mu_2 )
& = & \left\{ \left[ C( \Lambda ), E_{- \alpha_1}(\mu_1 ) \right]
                    E_{- \alpha_2} (\mu_2 )  \right.\\
&  & + E_{- \alpha_1} (\mu_1 ) \left[ C(\Lambda ),
                    E_{- \alpha_2}( \mu_2 ) \right] \\
&  & +  D(\mu_2 ) \left( \left[ C( \Lambda ) ,
                    E_{- \alpha_1}(\mu_1 ) \right] \right) \\
&  & +  \left. X^2 (\mu_1 , \mu_2 ) C(\Lambda )  \right\} \omega
\end{eqnarray*}
Inserting here the commutation relation (25) and pushing the arising
operators $h_\alpha , E_\alpha , (\alpha > 0 )$ to the right edge of
the expressions one obtains indeed - after evaluation of the action
of $h$ and $C$ on $\omega$ and after restriction to the pole
term - the r.h.s. of the OSBAE, Eq. (16).
\section*{Acknowledgements}
We thank M. Scheunert for patient instructions on Lie algebras.
One of us, (H.B.), thanks M. Karowski and R. Pogossian, the other
one (R.F.), thanks H. Grosse for discussions. H.B. thanks the
Physikalisches Institut in Bonn and R.F. thanks the
Yerevan Physics Institute for hospitality.


\begin{thebibliography}{99}
\bibitem[1]{} H.M. Babujian, in Proceed. XXIV Intern. Symp. Ahrenshoop,
 Zeuthen 1990, preprint YERPHI-1261 (4)-90.
\bibitem[2]{} M. Gaudin, J. Phys. (Paris) {\bf 37} (1976) 1087.
\bibitem[3]{} V.G. Knizhnik and A.B. Zamolodchikov,
 Nucl. Phys. {\bf B 247} (1984) 83.
\bibitem[4]{} H.M. Babujian, Off-Shell Bethe Ansatz Equation and N-Point
 Correlators in the SU(2) WZNW Theory, preprint Bonn-HE-93-22,
 Journ. of Physics A, in press.
\bibitem[5]{} L.D. Faddeev, E.K. Sklyanin and L.A. Takhtajan,
 Theor. Mat. Fis. {\bf 40} (1979) 194.
\bibitem[6]{} V.V. Schechtman and A.N. Varchenko, Invent. Math. {\bf 106}
 (1991) 139.
\bibitem[7]{} A. Matsuo, Commun. Math. Phys. {\bf 134} (1990) 65.
\bibitem[8]{} H. Awata, A. Tsuchiya and Y. Yamada, Nucl. Phys. {\bf B 365}
 (1991) 680.
\bibitem[9]{} H.J. de Vega, Int. J. Mod. Phys. {\bf A 4} (1989) 2371 and
references therein.
\bibitem[10]{} A.G. Ushveridze, preprint FTT-16 (Tbilisi, 1989).
\bibitem[11]{} B. Jur\v{c}o, J. Math. Phys. {\bf 30} (1989) 1289.
\bibitem[12]{} K. Hikami, P.P. Kulish and M. Wadati, J. Phys. Soc. Japan
{\bf 61} (1992) 3071.
\bibitem[13]{} H.M. Babujian and R. Flume, work in preparation.
\bibitem[14]{} A.A. Belavin and V.G. Drinfeld, Funct. Anal. Appl. {\bf 16}
 (1982) 159.
\end{thebibliography}
\end{document}